\documentclass[11pt]{article}
\usepackage[english]{babel}
\usepackage{lineno}
\usepackage[letterpaper,top=2cm,bottom=2cm,left=1.8cm,right=1.8cm,marginparwidth=1.75cm]{geometry}

\usepackage{amsmath}
\usepackage{graphicx}
\usepackage[colorlinks=true, allcolors=blue]{hyperref}

\title{Magnetization texture imprints produced by flux avalanches in  
ferromagnet/insulator/superconductor heterostructures}

\author{R.F. Lopes$^{1}$, M. A. Tumelero$^{1}$, C. I. L. de Araujo $^{2}$, A. M. H. de Andrade$^{1}$, F. Mesquita$^{1}$,\\ D. Carmo$^{3,4}$, F. Colauto$^{4}$, W.A. Ortiz$^{4}$, P. Pureur$^{1}$\\
\footnotesize{$^{1}$ Instituto de Física, Universidade Federal do Rio Grande do Sul, 91501-970 Porto Alegre, RS, Brazil.}\\
\footnotesize{$^{2}$ Departamento de Física, Laboratório de Spintrônica e Nanomagnetismo, Universidade Federal de Viçosa,} \\  
\footnotesize{36570-900 Viçosa, MG, Brazil.}\\
\footnotesize{$^{3}$ Laboratório Nacional de Luz Síncrotron, Centro Brasileiro de Pesquisa em Energia e Materiais, Campinas, 13053-970, SP, Brazil.} \\
\footnotesize{$^{4}$ Departamento de Física, Universidade Federal de São Carlos, 13565-905 São Carlos, SP, Brazil.} \\
}

\begin{document}
\maketitle

\begin{abstract}
The magnetic textures generated by a perpendicularly applied magnetic field at the ferromagnetic layer of $Co/Al_{2}O_{3}/Nb$ thin film heterostructures are investigated using magneto-optical imaging and micromagnetic simulations. It is observed that the stray field caused by flux avalanches in the superconducting layer prints out a non-trivial in-plane texture in the cobalt layer, which remains stable up to temperatures much above the Nb critical temperature. These textures mimic quite closely the dendritic thermomagnetic flux avalanches that penetrate the Nb layer from its edges. For low cobalt thickness, the filamentary magnetic textures occur in pairs with opposite polarity. The previous in-plane magnetization determines the relative location of the magnetic filaments with respect to the avalanche position. Micromagnetic simulations solving the Landau-Lifshitz-Gilbert equation confirm the interpretation given for the experimental findings.
\end{abstract}

\section{Introduction}
 
The rich variety of phenomena occurring at the interface of ferromagnets with a strong spin-orbit coupling material or a superconductor have been boosting the research on metallic heterostructures. The generation of magnetic textures in such interfaces is one of the most promising trends in the search for future magnetic and/or superconducting devices.
Particularly, due to improvements in thin film growth methods, nanostructure patterning technologies and cryogenics characterizations, the interplay between ferromagnetic (FM) and superconducting (SC) materials has become in recent years a prominent field of research in condensed matter physics, both from a fundamental and applied point of view.

In fundamental research, the possibility of observing correlated electron pairs with triplet configuration leading to superconductivity in FM/SC heterostructures has been a major goal in the area of superconducting spintronics\cite{Anwar_2016, Tkachov_2017}. Moreover, a proximity effect in a superconducting layer sandwiched between ferromagnetic insulating layers leads, for example, to the observation of infinite magnetoresistance induced by tuning the internal exchange field in the FM/SC interfaces \cite{Li_2013}. Also relevant are the investigations on superconducting magnonic effects in bilayers FM/SC \cite{Golovchanskiy_2018} and the search for Majorana modes using deformed skyrmions \cite{Garnier_2019}.

In the area of applications, some examples are the nanostructured junctions that have been employed in new devices as a field effect transistor, where superconductivity is manipulated by external electric fields applied at the junction interface \cite{Simoni_2018,Rocci_2021}, a Josephson phase battery \cite{Strambini_2020} or the high sensitive superconducting heat sensors for application in broad spectra radiation detection \cite{Heikkil_2018}.

The feasibility of imprinting topological protected magnetic textures in ferromagnetic layers showing perpendicular magnetic anisotropy (PMA) using arrays of micro-sized ferromagnetic disks have been proposed and demonstrated \cite{Sun_2013, Miao_2014, Loreto_2018}. Recently, the possibility for skyrmion and skyrmionium patterning and manipulation in PMA materials due to proximity of superconducting vortices have been investigated using numerical simulations and theoretical considerations \cite{DelValle_2015, Menezes_2019}. Such devices would enable fast and robust information record in magnetic racetracks \cite{Loreto_2019, Fert_2013}.

With the aim of investigating the presence and characteristics of texture imprints in magnetic layers, we have previously carried out a magnetotransport and magnetooptical study on a heterostructure formed by a ferromagnetic (FM) thin film separated by an insulator (I) layer from a superconducting (S) film where magnetic flux avalanches were generated \cite{Lopes_2017}. Our preliminary results demonstrated that the magnetic texture imprints on the ferromagnetic layer reproduce closely the dendritic configuration of the flux avalanches in the adjacent superconducting film, as also reported by Brisbois et al. \cite{Brisbois_2016}. The magnetic textures induced in our FM/I/S heterostructures remained stable far above the superconducting transition temperature. Furthermore, the entry of flux avalanches in the magnetic layer of our samples produces a characteristic noise in longitudinal magnetoresistance and Hall resistivity, which vanish at the onset of the critical field \cite{Lopes_2017}.

In order to deepen knowledge about the properties of such imprints of dendritic flux arrangements, in this work we report on further investigations on trilayer heterostructures with thin and thicker ferromagnetic films. The overall composition of the trilayers are $Nb(200nm) / Al_{2}O_{3}(15 nm) / Co(20nm)$ and $Nb(200nm) / Al_{2}O_{3}(15 nm) / Co(150nm)$. We show that the stray field generated by the flux avalanches in the superconducting layer imprints a robust magnetization texture on the adjacent FM layer, forming in-plane stripes that remain when the magnetic field is removed. A description of the mechanism that produces the magnetization textures is proposed. Micromagnetic simulations are presented to support the data analysis.

\section{Methods}

Thin film $Co/Al_{2}O_{3}/Nb$ heterostructures were prepared by magnetron sputtering deposition with the system Orion 8, manufactured by AJA International Inc. The following parameters were used: base pressure of $5 \times 10^{-8}$ Torr, argon working pressure of $2 \times 10^{-3}$ Torr under argon flux of 20 sccm. The Nb layers were deposited with DC power of 450 W, while the substrate temperature was kept at $500\: ^{0}\rm{C}$ in order to obtain a superconductor transition temperature as close as possible from that of bulk $Nb\:(T_{c} = 9.25 \: \rm{K})$ \cite{Feliciano_2016}. The $Al_{2}O_{3}$ layer was deposited with RF power of 200 W and the $Co$ layer with DC power of 150 W, both with substrate temperature around $20\: ^{0}\rm{C}$. A thin $Al_{2}O_{3}$ layer (10 nm) was deposited on top of the Co film as a protection against oxidation. The substrate used was a silicon wafer (100) oriented. The substrate was cleaned with piranha etching and hydrofluoridric acid (5 \%) solution prior to the deposition. The thicknesses of the layers were individually calibrated using X-ray reflectometry. Characterization of the individual layers was made with X-Ray diffraction. This analysis showed that the polycrystalline $Co$ and $Nb$ layers are hexagonal and cubic respectively, with the expected lattice parameters. The magneto-optical images were obtained in a setup based on the Faraday magneto-optical effect \cite{Johansen_1996}.  A $Bi$-substituted ferrite-garnet film with in-plane magnetization grown on a gadolinium gallium garnet (GGG) substrate was employed as a sensor \cite{Johansen_1996, Helseth_2002, Colauto_2007}. The sensor plate was laid down on the heterostructure surface. Then, the magnetic flux density distribution over the sample area was detected as an image in a polarized light microscope with crossed polarizers. A CCD camera with image acquisition frequency of 106 Hz was used to observe the magnetic flux penetration in the sample submitted to a perpendicular magnetic field \cite{colauto2020}. 

The micromagnetic simulations were carried out using the open source GPU based $MUMAX^{3}$ code, which numerically solve Landau-Lifshitz-Gilbert equation (LLG) \cite{Vansteenkiste_2011},

\begin{equation}
   \frac{\partial \vec{M}}{\partial {t}}= \gamma \Vec{H}{_{eff}}\times\Vec{M}+\frac{\alpha}{M_S}\times\frac{\partial\Vec{M}}{\partial{t}} 
\end{equation}

where $\gamma$ is the gyromagnetic ratio, $M_S$ is the saturation magnetization, $\alpha$ is the Gilbert damping coefficient and $H_{eff}$ is the effective field generated by the magnetocrystalline anisotropy, exchange and dipolar interactions. Finite discretization for the iterations following Eq. (1) was carried with cubic cells of $5  \times 5 \times 5 \: \rm{nm}^{3}$. Numerical parameters utilized for cobalt were the magnetic saturation $M_S = 1400 \times 10^{3} \rm{A/m}$, exchange constant $A_{ex}= 30 \times 10^{-12} \rm{J/m}$, damping $\alpha = 0.5$ and cubic anisotropy constant $520 \times 10^{3} \rm{J/ m^{3}}$.

\section{Results}

In Figure 1 are presented the magneto-optical images (MOI) for the heterostructure $Nb / Al_{2}O_{3} / Co$\,(20\,nm). The sample cross section and top view schemes are shown in the upper panels of Figs. 1(a) and (b). At 2.5 K, the perpendicularly applied magnetic field induces dendritic vortex avalanches in the superconducting film, a phenomenon previously described in references \cite{Lopes_2017, Brisbois_2016, ortiz_review, colauto2020}. The avalanche pattern obtained when $H$ = 46.5 Oe is shown in panel (ii) of Figure 1(a). The white dendritic lines are records of the flux avalanches penetrating from the edges of the $Nb$ layer. The width of the avalanche dendrites according to MOI in Fig. 1 ranges between 20 and 90 $\mu$m. These dendritic branches penetrate the area covered by the $Co$ film, whose position is indicated by the contour in orange dashed lines in panel (ii) of Figs.1(a) and 1(b). A magnified view of the flux avalanches is shown in the panel (iii) of Fig.1(a).

\begin{figure} [!h]
    \centering
    \includegraphics{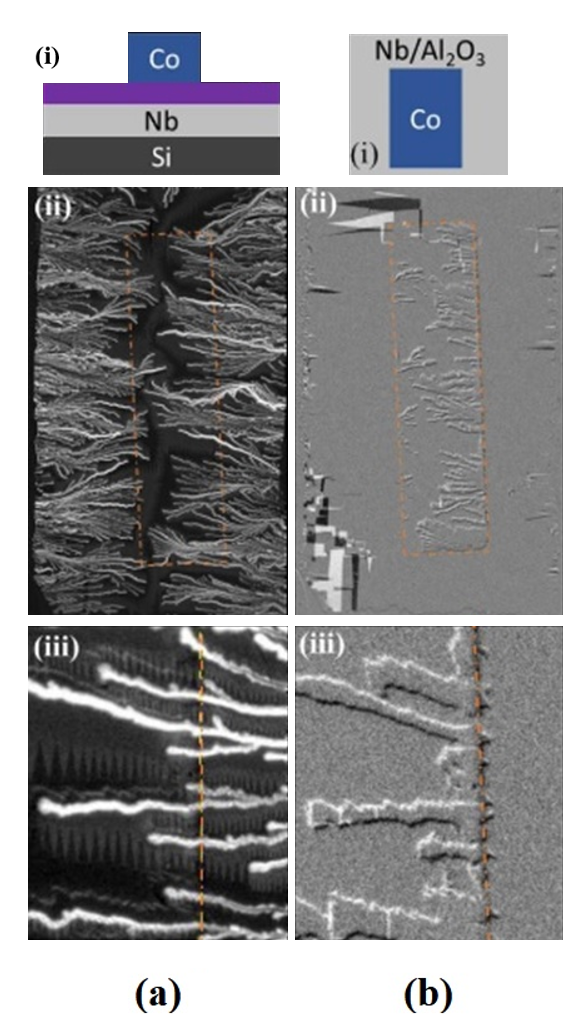}
    \caption{(a) Magneto-optical images obtained for the sample $Nb/Al_{2}O_{3}/Co$(20nm) at 2.5 K with $H$ = 46.5 Oe applied perpendicularly to the heterostructure; (i) the sample cross section scheme; (ii) overall view of the avalanche penetration in the samples, the orange dashed lines indicate the position of $Co$ overlayer; (iii) zoom view of the flux avalanches near the edge of the Co layer. (b) Magneto-optical images of the same sample recorded at 15 K and zero magnetic field; (i) top view scheme of the sample, (ii) overall MOI of the sample showing magnetization textures imprinted in the $Co$ layer by the superconducting flux avalanches; (iii) zoom view of the imprinted textures near the $Co/Nb$ edge.}
    \label{fig:Rovan_Fig.1}
\end{figure}

After capturing the MOI shown in Fig. 1(a), the magnetic field was removed and the temperature was increased to 15 K, safely above the critical temperature $T_c$  of niobium. As shown in the panel (ii) of Fig. 1(b), the dendritic flux penetrations disappear from the area occupied by the $Nb$ layer. However, magnetic texture imprints of the avalanches in the $Co$ layer (position marked by the contour in orange dashed lines) follow along the same path previously taken by the flux penetration in the superconducting layer and reproduce closely its dendritic shape. The panel (iii) of Fig.1(b) shows a zoom view of the magnetic texture corresponding the avalanches shown in panel (iii) of Fig.1(a).

Although there is resemblance between flux avalanches and magnetic textures in Fig.1, some details in the latter deserve closer examination. One observes that the textures are outlined by a bright edge on the top and by a dark edge on the bottom. In the used MOI setup, the bright lines indicate stray fields emerging from the film plane, while the dark lines indicate fields entering the plane. To further examine these features, we overlap the patterns shown in Figs.1(a) and 1(b) into one, as depicted schematically in panel (ii) of Fig.2(a). A zoom view of the enclosed area is presented in panel (iii) of Fig.2(a), which corresponds to the same region shown in panel (iii) of Figs.1(b) and (b). Results in Fig.2(a) were obtained at $T$ = 2.5 K by applying a magnetic field oriented upwards (emerging from the sample plane, or positive orientation), as illustrated in panel (i) of the figure. The thin black contours mostly seen in panel (iii) of Fig.2(a) delimits the regions where the avalanches are located. The blue and red lines represent the bright and dark filamentary cobalt textures at 15 K, respectively. The blue (bright) lines correspond to stray field field emerging from the plane, or positive orientation, while the red (dark) lines indicate stray field pointing downwards, or negative field orientation. Curiously, one observes that once the superconducting flux (field pointing upwards) is removed, the printed dendritic magnetic textures at coincident positions in the $Co$ layer correspond to a stray field entering the plane (red lines). Also interesting is that the bright (blue) lines of the cobalt texture (stray field leaving the plane) are placed outside (above in Fig. 2) the avalanche regions. The distance between such texture pairs is of the order of the avalanche width (20 – 90 $\mu$m). The same experiment and analysis were performed, but applying the external magnetic field in the opposite orientation. Results are shown in panel (b) of Fig. 2. Panel (i) of Fig.2(b) shows that the field is oriented downwards (negative field) in this case. The obtained MOI at 2.5 K and 15 K are schematized in panels (ii) and (iii) of Fig.2(b). Consistently with the first experiment, inside the avalanche dendrites the stray field imprinted in the $Co$ layer is opposite to that generated by the avalanches. One clearly observes in panel (iii) of Fig.2(b) that the blue and red lines shift downwards relative to the avalanche positions when the field orientation is reversed. The blue lines now lie within the avalanche regions whereas the red filaments remain off the avalanche dendrites. 

\begin{figure} 
\centering
\includegraphics{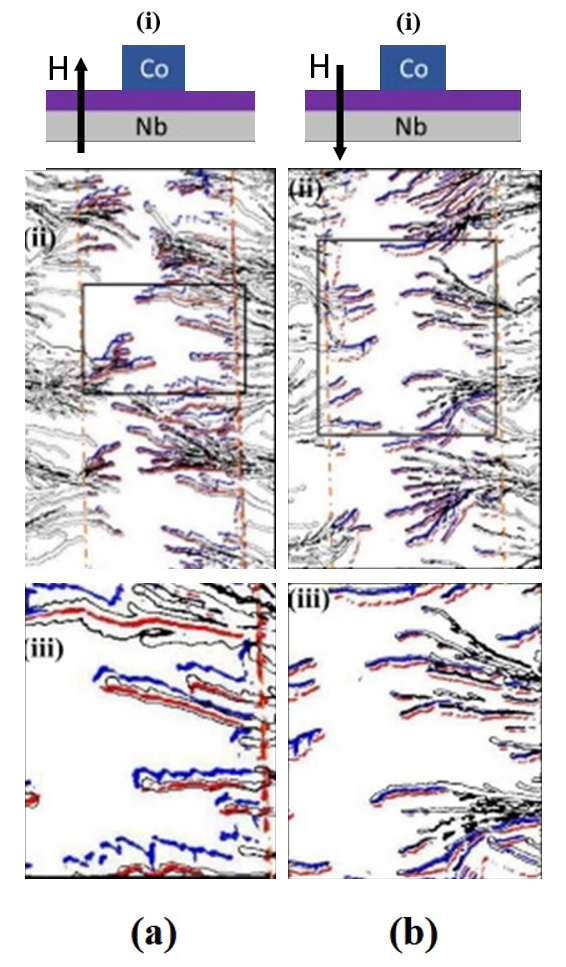}
\caption{Computer generated superposition of the magneto-optical images shown in Fig.1. Parts (a) and (b) of the figure correspond to opposite field orientations (i) Schematic representation of the field / sample geometry of the experiment. The field orientation is indicated by the black arrows. (ii) Superposition of images shown in panels (ii) of Fig.1, obtained in the superconducting and normal states. The contours of the avalanches are drawn with thin black lines; the blue and red lines represent the stray field of the magnetic textures leaving and entering the sample, respectively. (iii) Magnified views of the magnetization textures in the signaled areas in (ii).}
\end{figure}

We propose a model schematized in the cartoons of Figure 3 to account for the details of the texture imprinted in the cobalt layer by the superconducting flux avalanches. First of all, we observed that the $Co$ layer has a previous in-plane magnetization. MOI in Fig.3(a) shows the $Co$ layer before the application of the field that generates the flux avalanches. The dark and light edges of the $Co$ layer (stray field entering and leaving the film, respectively), reveal that an homogeneous magnetization is oriented approximately along a diagonal of the film, as indicated in the image. This easy in-plane axis for the magnetization comes from the combined contributions of the shape anisotropy of the film and the effect of local fields present during the deposition process. The abrupt entry of a dendritic flux avalanche deposits locally on the $Co$ film a significant amount of magnetic energy \cite{Altshuler_2004}, inducing a sharp and local reversal of the magnetization orientation, as illustrated schematically in panel (b) of Fig.3. Also represented in panel (b) are the filamentary textures generated by the stray field associated to the domain with reversed magnetization. Figure 3(c) shows a cross-section view of the heterostructure penetrated by the flux avalanche. The domain with reversed magnetization and the external stray fields associated with the filamentary magnetic textures are also represented. The depicted scenario allows one to understand why the stray field texture observed at coincidence with the avalanche position points downwards, being thus opposed to the upwards field corresponding to the superconducting flux. The entrance of the flux avalanches also reverses locally the orientation of the stray field at the edges of the $Co$ layer, when this one is parallel to the field generated by the avalanches. This effect is schematically illustrated by the red segments of the $Co / Nb$ border in Fig.3(b) and may be observed in Fig.2(a).

\begin{figure} [!h]
    \centering
    \includegraphics{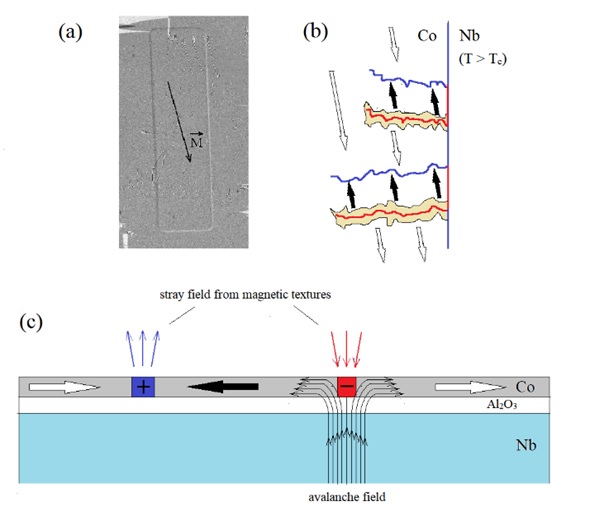}
    \caption{(a) Magneto-optical image showing the $Co$ film light and dark contours at $T$ = 2.5 K, observed previously to the application of the magnetic field which induces the flux avalanches. The orientation of the native homogeneous magnetization is shown. (b) Schematic top view of magnetization reversal produced by flux avalanches; the stray field filaments of the magnetization textures are represented by the blue and red lines (field leaving and entering the sample, respectively); the position of the avalanches is drawn in yellow encircled by thin black lines. The orientation of the native magnetization is indicated by white arrows, while the reversed domains are shown by black arrows.  (c) Cross-section view of the heterostructute showing the flux avalanche, the reversed magnetization domain (black arrow) and the up and downwards orientations of the stray-field associated with the filamentary magnetic textures at the upper surface of the sample. The magnetic circuit of the stray field closes in the air above the heterostructure surface, through the insulating  spacer and also through the London penetration depth of the Nb film.}
    \label{Rovan_Fig-3}
\end{figure}

The occurrence of a similar structure for the stray fields produced by the magnetic textures imprinted by flux avalanches in heterostructures $Py/SiO2/Nb$ is reported in \cite{Brisbois_2016}. However, in contrast to assumption made in that work, the filamentary textures observed in our samples cannot be attributed to usual Bloch or Néel domain walls. The widths of textures observed in our heterostuctures are in the range 4 – 10 $\mu$m, which are at least one order of magnitude larger than that of usual domain walls \cite{Suzuki_1969}. The spin configuration inside the magnetic dendrites of our sample is probably complex and we do not rule out the possibility of nucleation of skyrmion-type excitations or spin helices. 

We also studied the effect of the magnetic textures on the homogeneous flux penetration in the area covered by the $Co$ layer of the sample with $Co$(20 nm). These experiments were carried out in a region of the $H-T$ plane were flux avalanches do not occur \cite{Colauto_2008} and are presented in the Supplementary Material. 

Figure 4 shows MOI obtained for the heterostructure with thicker Co layer, $Nb / Al_{2}O_{3} / Co$(150 nm), in the superconducting and normal states. As in Fig. 3, results refer to the area covered by the $Co$ layer. Panel (a) shows a snapshot taken when $T$ = 2.5 K and $H$ = 46.5 Oe. The flux avalanches clearly invade the area covered by the $Co$ layer as also seen in the heterostructure with $Co$ (20 nm). However, some subtle diferences are noticeable: (i) There is a stronger contrast between the light intensity coming directly from the $Nb$ film and that coming through the $Co$ layer. The intensity ratio I($NbCo$)/I($Nb$) is approximately 0.65 for $Co$ (150 nm), whilst it amounts to 0.85 for $Co$ (20 nm). (ii) The width of the magnetic stripes in the $Co$ layer, which can be seen in panel (b), ranges in the interval 20 - 30 $\mu$m, much larger than those observed in the $Co$ (20 nm) sample. (iii) The double and split filamentary magnetic dendrites are not clearly visible, as 
shown by the analysis of image superposition in Figs. 4(c) and 4(d). Figure 4(c) corresponds to an applied field (and flux avalanche) emerging from the sample. Inside the region where the avalanches were previously located, the stray field of the imprinted filamentary magnetic texture (red colour) has opposite orientation, similarly to observations in the $Co$ (20 nm) sample. However the accompanying blue texture appears in most cases as a diffuse shading, suggesting that the closure of the reversed domain is rather widespread. Figure 4(d) shows the MOI registered when the field is applied downwards. Inside the contours defining the avalanche position (contours in black thin lines), the intense blue magnetic stripe corresponds to a stray field pointing upwards. In this case, the domain closures appears as a diffuse reddish shading next to the blue textures. The diffuse closure of the in-plane magnetization is probably related to the occurrence of frozen and small domains with out-of-plane magnetic moment. (iv) A fourth notable difference between the magnetic textures imprinted in the $Co$ (150 nm) sample with respect to that in $Co$ (20 nm) is the lower robustness against the temperature augmentation. Magneto-optical images taken at 105 K, 185 K and 235 K, shown in panels (e), (f) and (g) of Fig.4, respectively, reveal a progressive fading of the imprinted magnetic textures while the temperature increases. At $T$ = 235 K almost no texture can be observed. The relative weakness of the imprinted magnetic textures upon heating in the $Co$ (150 nm) is in line with our hypothesis that the closure of the reverted  in-plane magnetization in this sample occurs via frozen and small domains with out-of-plane orientation. These magnetic grains relax to the in-plane orientation due to thermal activation, thus erasing the textures. 
All of the mentioned differences between magnetization textures imprinted in the $Co$(150 nm) and the $Co$(20 nm) heterostuctures are direct consequences of the distinct $Co$ thickness. In particular, the greater $Co$ thickness makes the flux stray field spread over both in direction and intensity. The thickness of the $Co$ layer also plays a role in the shape anisotropy, which is smaller for the $Co$(150 nm) film, allowing the occurrence of an out-of-plane component of its magnetization. 

\begin{figure}
    \centering
    \includegraphics{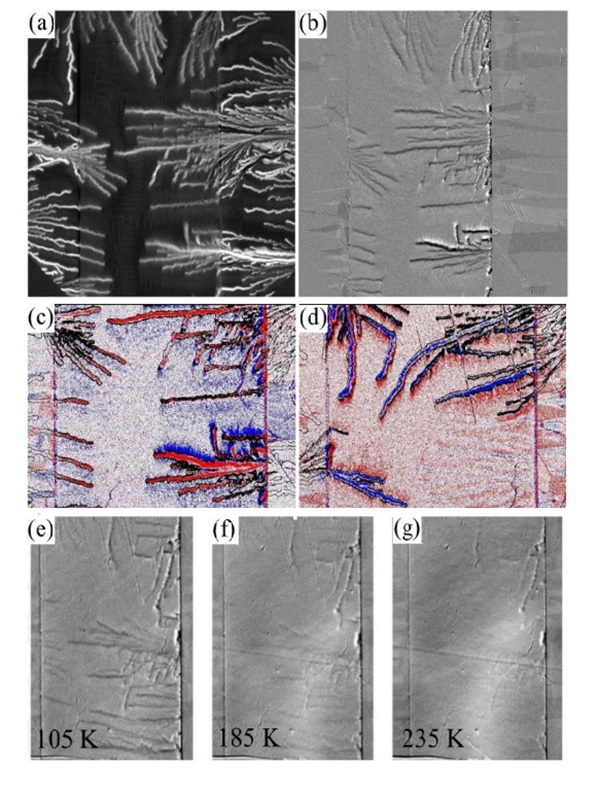}
    \caption{Magneto-optical images for sample $Nb$(200nm) / $Al_{2}O_{3}$(15nm) / $Co$(150nm). In (a) $T$ = 2.5K and $H$ = 46.5 Oe is oriented upwards. (b) Imprinted magnetic textures seen at $T$ = 15 K and  $H$ = 0. (c) Computer generated superposition of images in (a) and (b). The location of the avalanches is indicated by the thin black contours. The stray field related to the red filamentary texture points into the sample surface. (d) The same as in (c), but for the applied field pointing downwards. The blue lines correspond to stray field of the imprinted textures oriented upwards. MOI in (e) to (g) are taken in $H$ = 0 and $T$ = 105 K, 185 K and 235 K, respectively.}
    \label{Rovan_Fig-4}
\end{figure}

The micromagnetic simulations were carried out in a reduced cobalt film stripe with 1 $\mu$m × 1.5 $\mu$m × 20 nm (or 150 nm) dimension. Vertical periodic boundary conditions were adopted. A protocol with many steps of magnetic field ranging from 0.5 T down to zero was used to simulate a local avalanche generated by a field of 50 Oe applied perpendicularly to the $Nb$ film. Figure 5(a) shows the simulations for the sample $Co$ (20 nm), obtained for the downwards (negative) orientation of the stray field produced by the superconducting vortices.  An in-plane uniaxial anisotropy constant of 2 × $10^{5}$ J / $m^{3}$ was used in order to mimic the previous homogeneous magnetization existing in the $Co$ layer. The ground state magnetization observed after solving the LLG equation with the described protocol, represents the state remaining in the $Co$ film once the superconductivity was quenched and the stray field from vortices reduced to zero. One observes in panel (a) of Fig.5 that the field generated by the vortices revert the in-plane magnetization of the $Co$ layer in a region adjacent (at right-hand) to a line defined by the vortex cores (indicated by a dashed white line). The simulated magnetization profile is consistent with the experimental results presented in panel (iii) of Fig.2(a), where textures with opposite stray field orientation run parallel to each other.  In part (b) of Fig.5, simulations show that a similar scenario occurs when the field produced by the superconducting vortices is oriented upwards (positive field). It is interesting to notice that nearly the same stripe pattern occurs as for the downwards orientation, however, the stripe is now at the left-hand side of the avalanche line (indicated in white dashed), reproducing the observations in Fig.4. The simulations for the thicker $Co$ (150 nm) layer are shown in Fig.5(c). The 
rotation of the vectors is gradual and spread, which explains the blur seen in the magneto-optical images. Clearly, the smaller shape anisotropy favors magnetic imprints in the out-of-plane configuration. This fact explains why the imprints of the filamentary textures with out-of-plane magnetization are more intense in the magneto-optical images of figures 4(c) and 4(d).

\begin{figure} [!h]
    \centering
    \includegraphics[width=18cm]{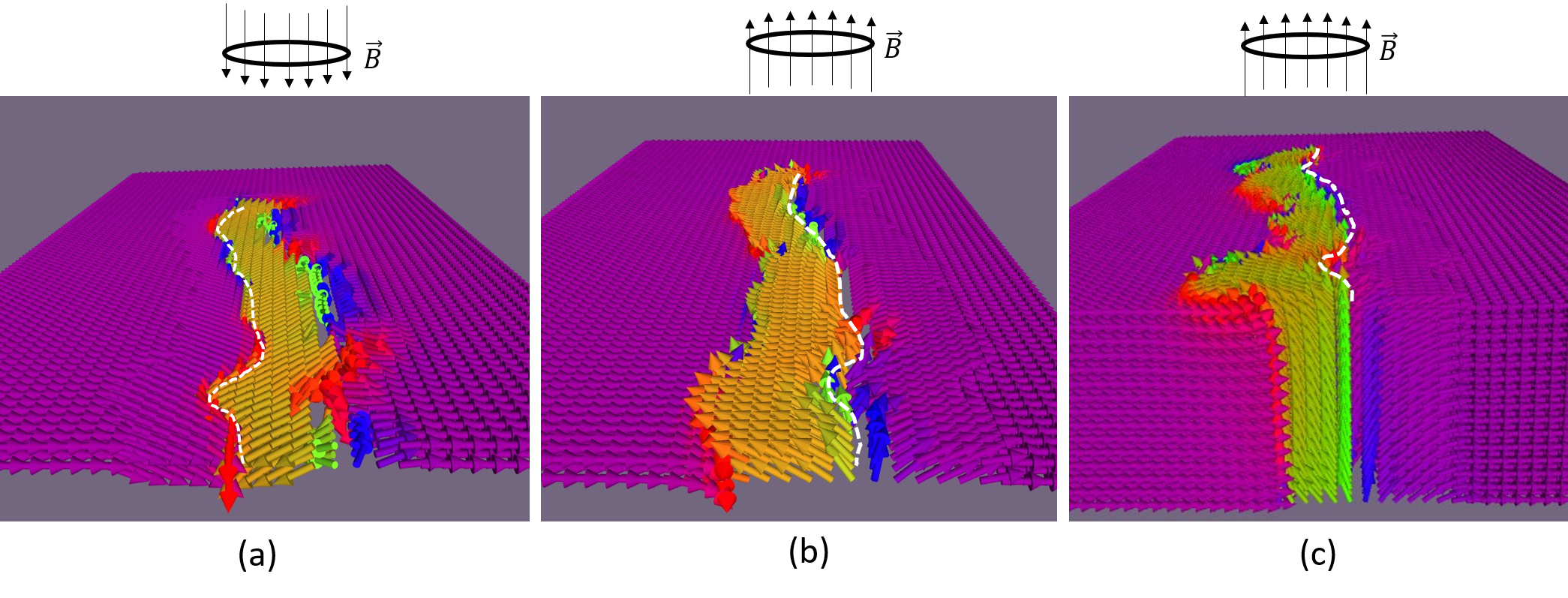}
    \caption{Magnetic ground state obtained from MUMAX3 simulations of (a) $Co$ (20 nm) thin film presenting the imprinted magnetization texture with downwards stray field (negative magnetic field) generated by the superconducting vortices whose position is signaled by a white dashed line. (b) The same as in (a) but for upwards vortex stray field (positive magnetic field). (c) Magnetization texture generated by upward vortex stray field in the $Co$ (150 nm) layer. The arrows indicate the magnetization directions and the color convention used is the following: red and blue arrows indicate magnetic moment mostly aligned along negative and positive z direction (into and out the sample plane), respectively. Purple indicate moments pointing to the right and parallel to the sample plane while orange indicate in-plane moments pointing to the left.}
    \label{Rovan_Fig-5}
\end{figure}

\section{Conclusion}

In summary, we investigated experimentally the imprints of magnetization textures on a cobalt layer by dendritic flux avalanches occurring in the adjacent niobium layer of a Superconductor/Insulator/Ferromagnet heterostructure. We observed that the dendritic magnetic textures generated at low temperatures may be stable up to room temperature for samples with small enough $Co$ thickness. The magnetic textures present a stray field configuration suggesting a local reversal of the previous in-plane $Co$ magnetization. Micromagnetic simulations performed using conventional parameters for ferromagnetic $Co$ films explain satisfactorily most the the experimental results and were helpful to understand the shift of the magnetization dendritic textures with respect to position of the flux avalanches that are observed in the magneto-optical images. We point out that the results presented here suggest that the field coming from the superconducting vortices can be applied in the magnetic recording technology and hopefully might be used to imprint skyrmions and other non-trivial textures in suitable magnetic materials.

\subsection*{Acknowlegedments}
We thank the Brazilian agencies Conselho Nacional de Ciencia e Tecnologia (CNPq) and Fundação de Amparo à Pequisa do Estado do Rio Grande do Sul (FAPERGS) for the financial support under the grants CNPQ 402915/2021-6, CNPQ 313809/2023,  PRONEX 16/0490-0. The authors also acknowledge the Laboratório de Conformação Nanométrica from Physics Institute for the sample preparation facilities. W.A. Ortiz and F. Colauto would like to acknowledge the São Paulo Research Foundation (FAPESP, Grant No. 2021/08781-8). The authors are also grateful to T. H. Johansen for inspiring discussions on the MOI technique and also on the physics content of the paper.

\bibliographystyle{unsrt}
\bibliography{ref_textures.bib}

\section{Supplementary Material}

The effect of magnetic textures imprinted in the $Co$ layer on the flux penetration was studied in a region of the $H-T$ plane were flux avalanches do not occur. Measurements were performed in the $Co$ (20 nm) heterostructure. Specifically, the sample temperature was fixed at $T$ = 8 K, and the applied field was varied in the range $H$ = 0 – 18 Oe. Magneto-optical images shown in figure S1 were first obtained when the $Co$ layer was demagnetized. Afterwards, MOI images were recorded at the same temperature and fields when the $Co$ layer was previously invaded by flux avalanches at low temperature then heated to $T$ = 15 K, so that superconductivity was quenched. In this second experiment, magnetic textures remain imprinted in the $Co$ layer. In the first case, the flux front progresses homogeneously through the area covered by the $Co$ layer upon field augmentation, roughly preserving the orientation parallel to the film edge. In the second experiment, the flux front is rather irregular. The nucleation of vortices along trajectories in the form of dendrites is clearly favored by the magnetic textures previously imprinted on the $Co$ layer, although no superconducting avalanche is expected to occur in the adopted experimental conditions. The textures whose stray field points opposite to the vortex orientation are more effective to guide the flux penetration, in accordance to previous studies on the interaction between vortices and domain walls of in-plane ferromagnets \cite{Milosevic_2004, Vestgarden_2007}.

\begin{figure}
    \centering
    \includegraphics[width=0.5\linewidth]{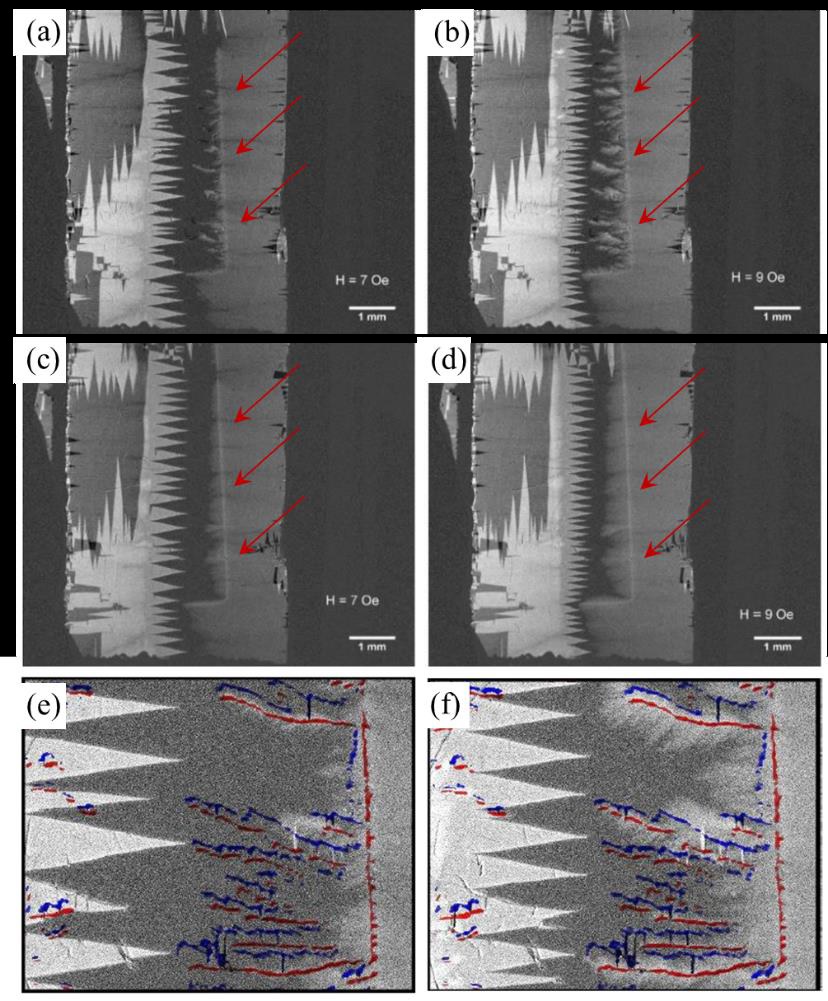}
    \caption{(a) MO image at 8 K and 7 Oe and (b) MO image at 8 K and 9 Oe. These two images were obtained after the following steps: (i) A field H = 46.5 Oe was applied perpendicularly to the sample at 2.5 K,  so that flux avalanches were produced (ii) The magnetic field was removed. (iii) The sample was heated up to 8 K. The red arrows indicate the orientation of the native in-plane magnetization of the $Co$ layer (c) MO image at 8 K and 7 Oe and (d) at 8 K and 9 Oe. These two MOI were obtained with the sample free from textures  . In panels (e) and (f), magnifications of the images shown in (a) and (b), respectively, were overlayed with corresponding magnetic textures registered in the Co layer after magnetization in 2.5 K and 46.5 Oe. The colour convention remains the same as in previous figures. The light and dark regions separated by saw-tooth boundaries in the MOI are due to magnetic domains in the ferrite-garnet indicator film.}
    \label{Rovan_Fig-S1}
\end{figure}

\end{document}